\documentstyle[twocolumn,aps]{revtex}

\begin{document}
\title{Magneto--Fluid Coupling --- Eruptive Events in the 
 Solar Corona}
\author{Shuichi Ohsaki$^{\,1}$, Nana L. Shatashvili$^{\,2\ *}$ \ and 
Zensho Yoshida$^{\,3}$ }
\address{
The University of Tokyo, Tokyo 113-0033, Japan}
\author{and\\
Swadesh M. Mahajan$^{\,4}$ }
\address{Institute for Fusion Studies, The University of Texas at Austin,
Austin,Texas 78712}

\maketitle

\begin{abstract}
By modelling the coronal structures by  "slowly" evolving Double--Beltrami 
two--fluid equilibria (created by the interaction of the magnetic and velocity 
fields), the conditions for catastrophic transformations of the original state 
are derived. It is shown that, at the transition, much of the magnetic energy of 
the original  state is converted to the the flow kinetic energy.  
 
\end{abstract}

{\it Subject headings:} \ Sun: flares --- Sun: corona ---- Sun: 
magnetic fields

\begin{flushleft}
\rule{3.4 in}{.007 in} \\
$^*$ \hskip 0.1cm {\small {\it Permanent address:} \ Plasma Physics Department, Tbilisi \\
\hskip 1.6cm  State University, Tbilisi 380028, Georgia} \\
$^{1}$  {\small E--mail: \ ohsaki@plasma.q.t.u-tokyo.ac.jp}\\
$^{2}$  {\small E--mail: \  nana@plasma.q.t.u-tokyo.ac.jp \\
\hskip 1.6cm nanas@iberiapac.ge}\\
$^{3}$  {\small E--mail: \ yoshida@k.u-tokyo.ac.jp}\\
$^{4}$  {\small E--mail: \ mahajan@mail.utexas.edu}\\

\end{flushleft}



\section{Introduction}
\label{sec:intro} The latest TRACE and SOHO/EIT observations have brought the 
solar corona into sharp focus. The observations reveal: 1) the structures that 
constitute the solar corona are in constant motion; they are full of fast--moving 
gases, and are heated primarily at their footpoints (base) very close to the 
solar surface. The heating occurs in few minutes in the first ten to twenty 
thousand kilometers above the surface, i.e, in a rather small fraction of the 
bright part of the anchored structure. In direct contradiction to the predictions 
of some theories, the heating is neither uniform (throughout the loops) nor does 
it happen preferentially near the top. A direct quote sums up the situation aptly: 
''Moreover, not only heat is deposited low down, but the gas is often actually 
thrust upward very rapidly. It does not merely `evaporate' into the coronal 
structures, it is often actually thrown up there. Exactly how that happens is 
still a puzzle'' \cite{bib:schrijver}, \ 2) the loops are composed of clusters of 
filamentary structures which are not, as believed before, static bodies supported 
by interior gas pressure and heated along their lengths \cite{bib:RTV}. They fill 
and drain so quickly that the gas in them must be moving nearly ballistically 
(see latest TRACE news and e.g. \cite{bib:schrijver}) along the substructures, 
rather than being ''quiescently heated''. From a detailed study of the loops with 
different characteristic parameters one concludes that the heating process is 
quite non--uniform \cite{bib:ANA}.
 
Transient brightenings, with their associated flows of cool and hot material,  
are also a very common phenomenon in the TRACE movies. These relatively fast 
(violent) happenings vary from small events in the quiet Sun to major flares  
in active regions; brightenings  which are more than $10^5$\,km apart often  
occur within the same exposure that typically lasts for 10 to 30\thinspace s 
\cite{bib:schrijver}. This kind of a coincidence in the events at distant 
locations is suggestive of fast particle beams propagation along separate magnetic 
loops which come together at the flaring site. The flaring sites are
generally assumed to be reconnection sites although observations have 
not establish a causal connection: ''Direct evidence for reconnection in 
flares is difficult to find, despite the fact that it is thought to be the 
primary process behind flares'' \cite{bib:schrijver}. It is remarkable that 
often the post--flare loop systems begin to glow at the TRACE EUV wavelengths 
without substantial distortion: reconnection that probably took place appears 
to be (largely) completed by the time the loops are detected.

These observations pose a new challenge for the theories of quiescent as well 
as not so quiescent coronal structures and  events. In this paper we examine 
the conjecture that the formation and primary heating of the coronal structures 
as well as the more violent events (possibly flares, erupting prominences and 
coronal mass ejections (CMEs)) are the expressions of different aspects of the 
same general global  dynamics that operates in a given coronal region  ~\cite{bib:MMNS}. 
It is stipulated  that the coronal structures are created from the evolution and 
re--organization of a relatively cold plasma flow emerging from the subcoronal 
region and interacting with the ambient solar magnetic field. The plasma flows, 
the source of both the particles and energy (part of which is converted to heat), 
in their interaction with the magnetic field, also become dynamic determinants of 
a wide variety of plasma states; it is likely that this interaction may be the cause 
of the immense diversity of the observed coronal structures 
~\cite{bib:MMNS,bib:MY-1,bib:MY-2}. Preliminary results from this magneto--fluid 
approach  reproduce several of the salient observational features of the typical 
loops: the structure creation and primary heating are simultaneous -- the heating 
takes place  (by the viscous dissipation of the flow kinetic energy) in a few minutes, 
is quite non-uniform, and the base of the hot structure is hotter than the rest.
 
We plan to extend the scope of the magneto-fluid theory beyond
the creation of the semi-quiescent  coronal structures by seeking 
answers to the following: a) can the basic framework of this model predict the 
possibility of, and the pathways for the occurrence of sudden, eruptive, and 
catstrophic events (such as flares, eruptive prominences, CMEs) in the solar 
atmosphere, b) does the eventual fate, possibly catstrophic reorganization, of 
a given structure lie in the very conditions of its birth, c) is it possible to 
identify the range and relative values of identifiable physical quantities that 
make a given structure prone to eruption (flaring), d) will an eruption be 
the result of the conversion of excess magnetic energy into heat and bulk 
plasma motion as is generally believed to happen in the solar atmosphere 
~\cite{bib:sakurai}-\cite{bib:sturrok}\ ?

We begin by identifying the quasi--equilibrium state of a typical coronal sturcture 
with a slowly changing  Double--Beltrami (DB) state (one of the simplest, non--trivial 
magnetofluid equilibrium). The slow changes may be due to changes in the sun which 
affect the local magnetic fields, the interaction of various nearby structures, or 
disturbances in the solar atmosphere. The parameter change is  assumed to be 
sufficiently slow that, at each stage, the system can find its local DB equilibrium 
(adiabatic evolution). The slow evolution must conserve the dynamical invariants: 
the helicity $h_{1}$, the generalized helicity $h_{2}$, and the total (magnetic plus 
the fluid) energy $E$. The problem of predicting sudden events (e.g.
catastrophic eruption) then reduces to finding the range, if any, in 
which the slowly evolving structure may suffer a loss of equilibrium. The signature 
of the loss of equilibrium is quite easy to identify for the DB states.
The transition may occur in one of the following two 
ways: 1) when the roots of the quadratic equation, determining the 
length scales for the field variation, go from being real to complex 
(implying change in the topology of the magnetic and the velocity fields --- 
boundary separating the paramagnetic from the diamagnetic), or 2) the amplitude of 
either of the two states ceases to be real. For our current problem, the sudden 
change is likely to follow the second route.

By analysing a simple analytically tractable model, we find  
affirmative answers to all the four questions we posed. We show that 
the invariants $h_1, h_2$ , and $E$ , which label and alongwith the 
initial and bounbary conditions determine the original state, hold the key to the 
eventual fate of a structure. If for a given equilibrium sequence, the  total 
energy $E$ is larger than some critical value (given in terms of $h_1$, and $h_2$), the 
catastrophic loss of equilibrium could certainly occur. The trigger for the 
equilibrium loss could come, for instance, from nearby  structures getting close 
to each other with an increase in their interaction energy. The catastrophy pushes 
a DB state to relax to a minimum energy single Beltrami field. For coronal 
structures, the transition  transfers almost all the short--scale magnetic energy to 
the flow energy.

\section{Model}

\label{sec:Model} 

Within the framework of our approach, there are two distinct scenerios for eruptive 
events : a) when a "slowly" evolving structure finds itself in a state of no 
equilibrium,  and b) when the process of creating a long--lived hot structure 
is prematurally aborted; the flow shrinks/distorts the structure which suddenly 
shines and/or releases energy or ejects particles. The latter mechanism requires 
a detailed time--dependent treatment and is not the subject matter of this paper. 
The following semi--equilibrium, collisionless magneto--fluid treatment pertains 
only to the former case. 

\bigskip

A given structure is supposed to corresspond to the equilibrium solutions of the 
two--fluid system 
\begin{equation}
\frac{\partial}{\partial t} \bbox{\omega}_j - 
\nabla\times(\bbox{U}_j\times%
\bbox{\omega}_j) = 0 \quad (j = 1,2)  \label{Vdynamics1}
\end{equation}
written in terms of a pair of generalized vorticities 
$\bbox{\omega}_1 = \bbox{B}, \ \bbox{\omega}_2 = \bbox{B} + 
\nabla\times \bbox{V}$,
and  effective flows 
$\bbox{U}_1 = \bbox{V} - \nabla\times \bbox{B}, \ \bbox{U}_2 = 
\bbox{V}$, with the following normalizations: the magnetic field 
$\mbox{\boldmath $B$}$ to an appropriate measure of the magnetic field $B_0$, 
the fluid velocity $\mbox{\boldmath $V$}$ to the corresponding Alfv\'en speed, 
and the distances to the collisiolless ion skin depth $l_i$.
  
The simplest and perhaps the most fundamental equilibrium solution to 
(\ref{Vdynamics1}) is given by the ``Beltrami conditions'', which imply the 
alignment of the vorticities and the corresponding flows 
($\mbox{\boldmath $\omega$}_j$//$\mbox{\boldmath $U$}_j$),
\begin{eqnarray}
& & \mbox{\boldmath $B$} = a (\mbox{\boldmath $V$} - \nabla\times%
\mbox{\boldmath $B$}),  \label{eq:D-Beltrami1} \\
& & \mbox{\boldmath $B$} + \nabla\times\mbox{\boldmath $V$} = b %
\mbox{\boldmath $V$},  \label{eq:D-Beltrami2}
\end{eqnarray}
with $a,\,b=const$. We have used constant density assumption for
simplicity - extension to varying density is straightforward 
\cite{bib:MMNS}. Equations (\ref{eq:D-Beltrami1}) and (\ref{eq:D-Beltrami2}) 
combine to yield: 
\begin{equation}
({\rm curl} - \lambda_+)({\rm curl} - \lambda_-) \mbox{\boldmath $B$} = 0, 
\label{eq:D-Beltrami3} \\
\end{equation}
where $\tilde{a}=1/a$ , $\nabla \times $ = ``curl'', and  
\begin{equation}
\lambda_{\pm} = \frac{1}{2} \left[(b-\tilde{a}) \pm 
\sqrt{(b+\tilde{a})^2 - 4}\,\right]. \label{eq:lambda_pm}
\end{equation}
  
For sub--alfvenic flows (the flows we generally encounter in 
the solar atmosphere), the length scales ($\lambda_{\pm}^{-1}$) are 
quite disparate. We assume $\lambda _{+}^{-1} >> \lambda_{-}^{-1}$ without loss 
of generality. The general solution to the ``double Beltrami equations'' 
(\ref {eq:D-Beltrami3}) is a linear combination of the single Beltrami fields 
$\mbox{\boldmath $G$}_\pm$ satisfying $({\rm curl} - \lambda) \mbox{\boldmath $G$}=0$. 
Thus, for arbitrary constants $C_\pm$, the sum 
\begin{equation}
\mbox{\boldmath $B$} = C_+ \mbox{\boldmath $G$}_+ + C_- \mbox{\boldmath $G$}_-  
\label{eq:solutionB}
\end{equation}
solves (\ref{eq:D-Beltrami3}), and the corresponding flow is given by 
$\mbox{\boldmath $V$} = \left(\lambda_+ + \tilde{a}\right)C_+ 
\mbox{\boldmath $G$}_+ + \left(\lambda_- + \tilde{a}\right)C_- \mbox{\boldmath $G$}_- $.
 
The DB field encompasses a wide class of steady states of mathematical physics -- 
from the force--free paramagnetic field to the fully diamagnetic field.  The 
Beltrami conditions  also demand ``generalized Bernoulli conditions'' which 
allow pressure confinement when an appropriate flow is driven \cite{bib:MY-1} 
(and references therein).

\bigskip
 
The DB field is characterized by four parameters: $\lambda_+, \, 
\lambda_-$ (eigenvalues), and $C_+, \,C_-$ (amplitudes). The three invariants 
\cite{stein}: the helicity $h_{1}$, the generalized helicity $h_{2}$,  
\begin{eqnarray}
h_{1} &=&\frac{1}{2}\int(\mbox{\boldmath $A$}\cdot \mbox{\boldmath
$B$})\,{\rm d}{\bf r} ,  \label{eq:h1-0} \\
h_{2} &=&\frac{1}{2}\int(\mbox{\boldmath $A$}+\mbox{\boldmath $V$}) 
\cdot (\mbox{\boldmath $B$}+\nabla \times \mbox{\boldmath $V$})\,{\rm 
d}{\bf r} ,  \label{eq:h2-0}
\end{eqnarray}
($\mbox{\boldmath $A$}$ is the vector potential), and the total energy 
\begin{equation}
E=\frac{1}{2}\int(\mbox{\boldmath $B$}^{2}+\mbox{\boldmath $V$}^{2})
\,{\rm d}{\bf r}   \label{eq:E-0}
\end{equation}
will provide three algebraic relations connecting them ~\cite{bib:ymoin}. To predict 
the possibility 
of an eruptive event, interpretted as the termination of an equilibrium sequence 
(for solar flares, this kind of an approach, albeit in different contexts, has been 
followed in numerous investigations, (see e.g. \cite{bib:forbes1}-\cite{bib:kusano1} and
references therein), we analytically investigate this system using the macro--scale 
($\lambda _{+}^{-1}$) of the closed  strcuture as a control parameter. This choice 
is physically sensible and is motivated by observations because in the process of
structure--structure interactions, ''initial'' shapes  do undergo
deformations/distortions with rates strongly dependent on the 
initial and boundary conditions.

For simplicity we explicitly work out the system in a Cartesian cube of length $L$. 
We take $\mbox{\boldmath $G$}_{\pm}$ to be the simple 2-D Beltrami ABC field 
\cite{bib:arnold},
\begin{equation}
\mbox{\boldmath $G$}_{\pm }=g_{x\pm }\left( 
\begin{array}{c}
0 \\ 
\sin {\lambda _{\pm }x} \\ 
\cos {\lambda _{\pm }x}
\end{array}
\right) +g_{y\pm }\left( 
\begin{array}{c}
\cos {\lambda _{\pm }y} \\ 
0 \\ 
\sin {\lambda _{\pm }y}
\end{array}
\right) ,  \label{eq:ABC-1}
\end{equation}
with the normalization $(g_{x\pm })^{2}+(g_{y\pm })^{2}=1$. For real 
${\lambda _{\pm }}$, (\ref{eq:ABC-1}) represents an arcade--magnetic field structure  
resembling interacting coronal loops [in Fig.~1]. Assuming 
$L=n_{+}(2\pi /\lambda _{+})=n_{-}(2\pi /\lambda _{-})$ ($n_{\pm }$ are integers), 
$\mbox{\boldmath $G$}_{\pm }$ satisfy the following relations: 
$\int\mbox{\boldmath $G$}_{\pm }^{2}{\rm d}{\bf r} =
L^{2},\ \int\mbox{\boldmath $G$}_{+}\cdot \mbox{\boldmath $G$}_{-}%
{\rm d}{\bf r} =0$, \ where $\int\,{\rm d}{\bf r}
=\int_{0}^{L}\int_{0}^{L}\,{\rm d}x{\rm d}y$.

The invariants can now be readily evaluated and the results can be displayed in 
several equivalent forms. We find the following three equations to be the  
most convenient for further analysis ($h_{2}=h_{1}+\tilde{h}_{2}$, \ 
$\tilde{h}_{2}=bE-\lambda _{+}\lambda _{-}h_{1}$):
\begin{eqnarray}
\tilde{h}_{2} &=&\frac{E}{2}\left[ ( \lambda _{+}+\lambda _{-})
\pm \sqrt{( \lambda _{+}-\lambda _{-}) ^{2}+4}\,\right]  \nonumber \\
&&\quad \quad -\lambda _{+}\lambda _{-}h_{1},\label{eq:h2-1}\\
C_{+}^{2} &=&D^{-1}\left\{ E-\left[ 1+( \lambda 
_{-}+\tilde{a})^{2}\right] 
\lambda _{-}h_{1}\right\} \lambda _{+},\label{eq:C1-2} \\
C_{-}^{2} &=&- D^{-1}\left\{ E-\left[ 1+( \lambda 
_{+}+\tilde{a})^{2}\right] 
\lambda _{+}h_{1}\right\} \lambda _{-},\label{eq:C2-2}
\end{eqnarray}
where we have removed the common factor $L^2/2$, and
\begin{eqnarray}
D &=&\left[ 1+(\lambda _{+}+\tilde{a}) ^{2}\right] \lambda
_{+}-\left[ 1+(\lambda _{-}+\tilde{a}) ^{2}\right] \lambda _{-} \nonumber \\
&=&(\lambda _{+}-\lambda _{-}) b(b+\tilde{a}) .  \label{eq:assume-ab1}
\end{eqnarray}
For given $h_{1}$, $E$, $\tilde{h}_{2}$ ($h_{2}$) and $\lambda _{+}$
(control parameter), we can solve the preceding system to determine  
the physical quantities $\lambda _{-}$, and $C_{\pm}$ which must all 
remain real for an equilibrium. Before we give an analytic derivation 
for the bifurcation conditions (leading to loss of  equilibrium), 
we display in Fig.2 the plots of $\lambda _{-}$ and $C_{\pm }$ as 
functions of $\lambda _{+}$ for two distinct sets for the values of the invariants: 
we choose $h_{1}=1$, $h_{2}=1.5$, $E=0.4$  for Fig.2(a), and $h_{1}=1$, $h_{2}=1.5$, 
$E=1.3$  for Fig.2(b) (dashed lines correspond to the region of imaginary $C_{-}$). 
We find that the behavior of the solution changes drastically with $E$. For the 
parameters of Fig.2(a), $\lambda _{-}$ and $C_{\pm }$ remain real and 
change continuously with varying $\lambda _{+}$ implying that as the 
macroscopic scale of the structure ($1/\lambda _{+}$) changes continuously, 
the equilibrium expressed by (\ref{eq:ABC-1}) persists -- there is no catastrophic 
or qualitative change. For Fig.2(b) with $E$ changing from 0.4 to 1.3 (with same 
$h_{1}, \, h_{2}$) we arrive at a fundamentally different situation; when 
$\lambda _{+}$ exceeds a critical value $\lambda _{+}^{{\rm crit}}$, i.e, 
the macro--scale becomes smaller than a critical size, the physical determinants 
of the equilibrium cease to be real; the sequence of equilibria is terminated.
 
The condition for catasrophe turns out to be a constraint involving 
$h_{1}, \,h_{2}$ and $E$, which will allow the vanishing of 
$C_{-}^{2}$ for positive $C_{+}^{2}$, and real $\lambda_{\pm}$.
It is straightforward to show that the system has a critical 
point if
\begin{equation}
E^{2}\geq E_{c}^{2}=4\left( h_{1}\pm \sqrt{h_{1}h_{2}}\right) ^{2}
\label{eq:catascondition-1}
\end{equation}
and the critical $\lambda$ is determined by a simultaneous solution 
of (11) and $E-\left[ 1+( \lambda _{+}+\tilde{a})^{2}\right] 
\lambda _{+}h_{1}=0 $
\ giving:
\begin{equation}
\lambda_{+}^{{\rm crit}}=\frac{1}{2h_1}\,\left( E \pm \sqrt{E^2 - E_c^2}\right) \ .
\end{equation}

Thus, for  $E>E_c$ (determined by helicities $h_{1}$ and $h_{2}$),
when the macroscopic size of a structure shrinks below a critical value,
it can go through a severe reorganization.

At the critical point, an expected but most remarkable transition occurs. Using 
the value of $\lambda_{+}^{{\rm crit}}$, we find from
equation(\ref{eq:C2-2}) that the coefficient $C_{-}$\,, which measures the strength 
of the short scale fields, identically 
vanishes, and the equilibrium changes from Double Beltrami to a single Beltrami 
state defined by  $\lambda_{+}=\lambda_{+}^{{\rm crit}}$, i.e., 
$\mbox{\boldmath $B$}=C_{+}\mbox{\boldmath $G$}_{+}$ \ ($\nabla \times 
\mbox{\boldmath $B$}=\lambda _{+}\mbox{\boldmath $B$}$) \ with $\mbox{\boldmath $V$}$ 
parallel to $\mbox{\boldmath $B$}$. The transition leads to a magnetically more 
relaxed state with the magnetic energy reaching its minimum with appropriate gain 
in the flow kinetic energy (see Fig.\ref{fig:solar}).

\section{Conclusions}

\label{sec:conclusions}

By modelling quasi--equilibrium, slowly evolving coronal structures as a sequence 
of Double--Beltrami magnetofiuid states in which the magnetic and the velocity field 
are self--consistently coupled, we have shown the possibility of, and derived the 
conditions for catastrophic changes leading to a fundamental transformation of 
the initial state. The critical condition comes out as an inequality involving 
three invariants of the collisionless  magnetofluid dynamics. When the total energy  
exceeds a critical energy the DB equilibrium suddenly relaxes to a single Beltrami 
state corresponding to the large macroscopic size. All of the short--scale magnetic
energy is lost having been transformed 
to the flow energy and partly to heat via the viscous dissipation of 
the flow energy.

This general mechanism in which the flows (and their interactions with the magnetic 
field) play an essential role could certainly help in advancing our understanding 
of a variety of sudden (violent) events in the solar atmosphere like the flares, 
the erupting prominences, and the coronal mass ejctions. The connection of flows 
with eruptive events is rather direct: it depends on their ability to deform (in 
specific cases distort) the ambient magnetic field lines to temporarily stretch 
(shrink, destroy) the closed field lines so that the flow can escape the local 
region with a considerable increase in kinetic energy in the form of heat/bulk motion.

\bigskip

For SMM  this study was supported by US Department of Energy Contract 
No.DE-FG03-96ER-54366. The work of ZY was supported by Grant--in--Aid for Scientific 
Research from the Japanese Ministry of Education, Science and Culture No.09308011. 
Work of NLS was partially supported by the Joint INTAS--Georgian Grant No.52.
Authors thank Abdus Salam International Centre for Theoretical Physics, Trieste, Italy, 
where this work was started.


\begin{figure}[tbp]
\caption{Magnetic field line structure of a 2-D ABC map resembling 
coronal arcades.}
\label{fig:corona}
\end{figure}

\begin{figure}[tbp]
\caption{(a). Plots of $\lambda_-$ and $C_{\pm}$ versus $\lambda_+$ for $E = 0.4 
< E_c \simeq 0.45$, the critical energy. No catastrophe. \ 
(b). Plots of $\lambda_-$ and $C_{\pm}$ versus $\lambda_+$ for $E = 1.3 > E_c$ . 
There is a critical point at $\lambda_+ \simeq 0.041$. }
\label{fig:behavior}
\end{figure}

\begin{figure}[tbp]
\caption{Plots for $\lambda_-$, the magnetic and  the flow energies 
versus $\lambda_{+}$ for the catastrophe--prone set $h_{1}=1$, $h_{2}=1.007$ and 
$E=1.3>E_c=7\cdot 10^{-3}$. The scale lengths are highly separated $\lambda_{+}<<\lambda_{-}$. The 
initial choice makes $C_{+}\sim O\left( \lambda _{+}/\lambda_{-}\right)\ll 1$ 
and $C_{-}\sim O\left( \lambda _{-}/\lambda _{-}\right) \sim 1$ from (\ref{eq:C1-2}) 
and (\ref{eq:C2-2}). If any interaction increases $\lambda _{+}$ (the size of the 
structure shrinks)  the critical point ($\lambda _{+}=\lambda _{+}^{{\rm crit}}$) 
will be reached at which $C_{-}$ is zero. The magnetic field energy ($\propto C_{+}^{2}
+C_{-}^{2}$) decreases to a very small value since $C_{+}^{2}\ll 1$. Since the total 
energy is conserved, almost  all the initial magnetic energy is transfered to the flow 
causing an eruption. Notice that for coronal plasma, the skin depth 
$l_{i}$ is small $\sim 100{\rm cm}$ ($l_{i}/\lambda _{+}\sim 10^{3}{\rm 
km}$), for a typical density of $\sim 10^9\,{\rm cm}^{-3}$.
A word of caution is necessary -- as we approach the  critical point, 
the quasiequilibrium considerations are just an indicator of what is 
happening -- they must be replaced by a full time--dependent treatment 
to capture the dynamics; the changes are no longer slow.} 
\label{fig:solar}
\end{figure}






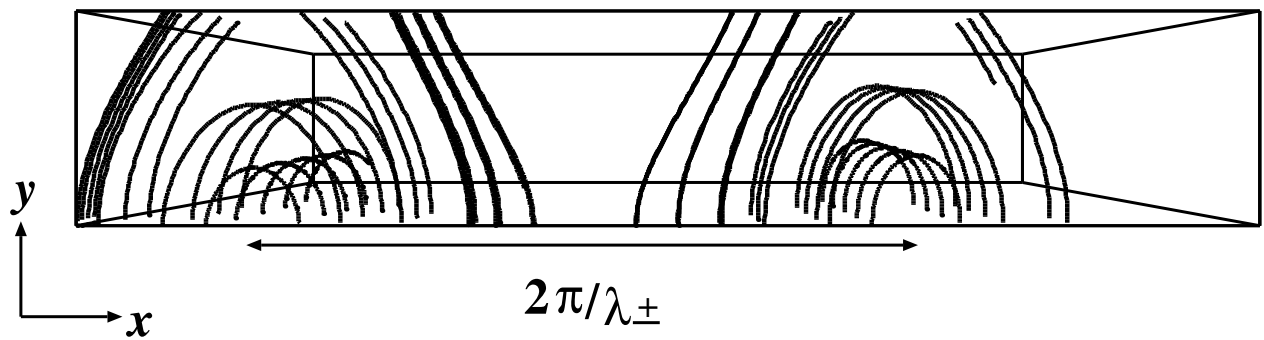
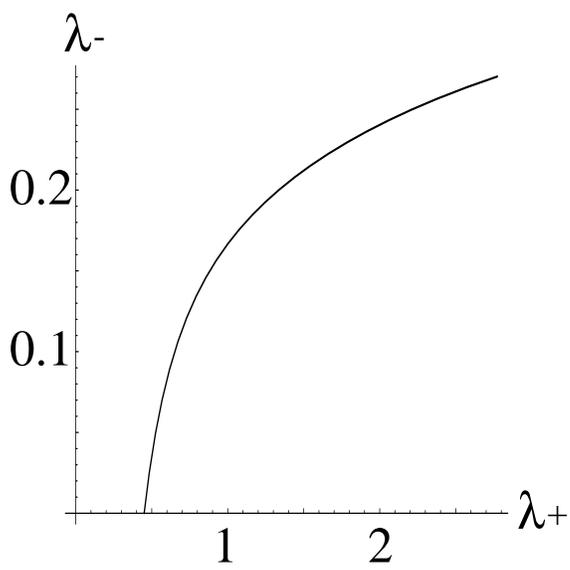
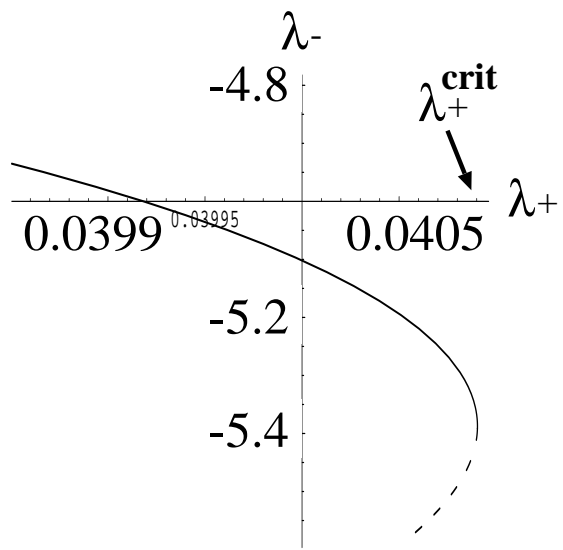
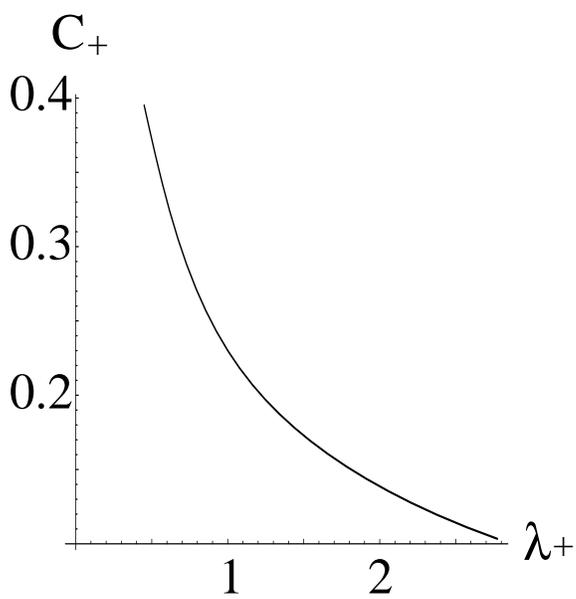
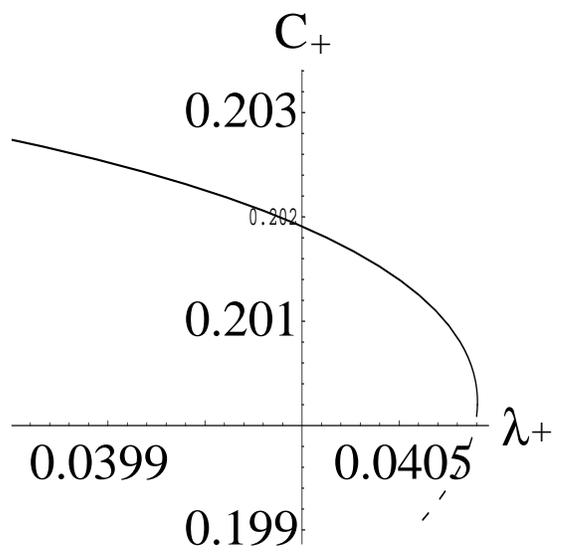
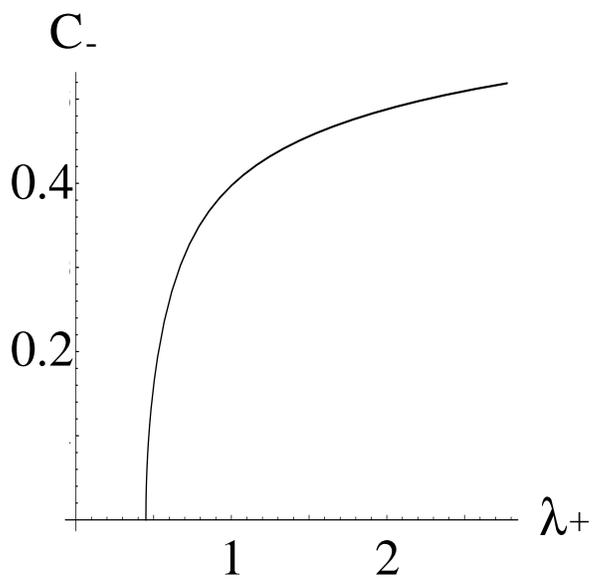
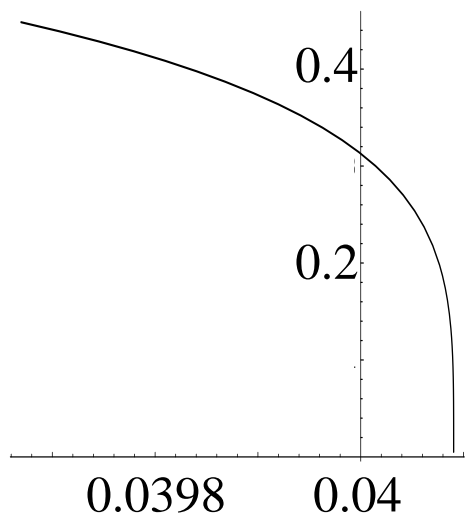
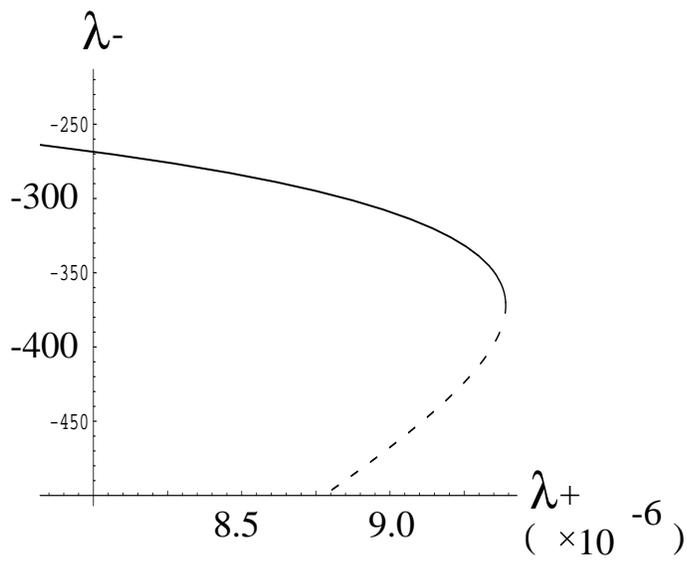
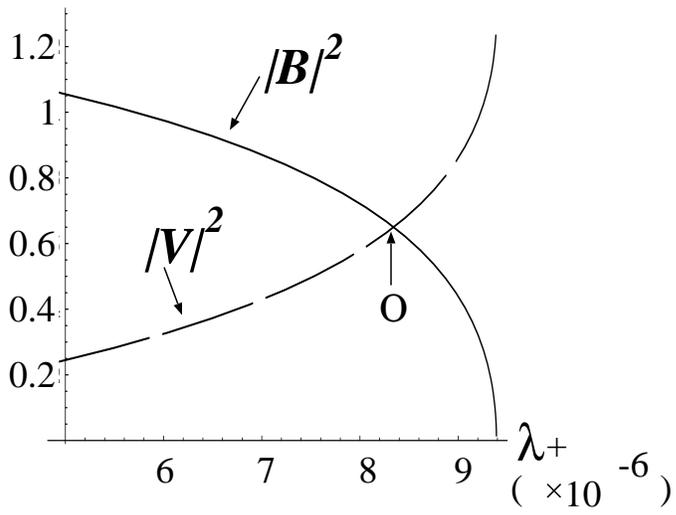
\end{document}